\documentclass[%
 reprint,
superscriptaddress,
longbibliography,
 amsmath,amssymb,
  pre
]{revtex4-2}
\usepackage{graphicx} 

\usepackage{textgreek} 
\usepackage{physics}
\usepackage[colorlinks=true,citecolor=blue,linkcolor=red,linktocpage=true]{hyperref}
\usepackage{xcolor}

\newcommand{\rev}[1]{\textcolor{black}{#1}}

\begin{document}
\preprint{APS/123-QED}

\title{Topology of the simplest gene switch}
\author{Aleksandra~Nelson}
\affiliation{Center for Theoretical Biological Physics, Rice University, Houston TX, 77005, USA}
\author{Peter Wolynes}
\affiliation{Center for Theoretical Biological Physics, Rice University, Houston TX, 77005, USA}
\affiliation{Department of Chemistry, Rice University, Houston TX, 77005, USA}
\affiliation{%
 Department of Physics and Astronomy, Rice University, Houston TX, 77005, USA
}
\affiliation{Department of Biosciences, Rice University, Houston TX, 77005, USA}
\author{Evelyn~Tang}
\affiliation{Center for Theoretical Biological Physics, Rice University, Houston TX, 77005, USA}
\affiliation{%
 Department of Physics and Astronomy, Rice University, Houston TX, 77005, USA
}
\date{\today}

\begin{abstract}
    Complex gene regulatory networks often display emergent simple behavior. Sometimes this simplicity can be traced to a nearly equivalent energy landscape, but not always. Here we show how a topological theory for stochastic and biochemical networks can predict phase transitions between dynamical regimes, where the simplest landscape paradigm would fail. We demonstrate the utility of this topological approach for a simple gene network, revealing a new oscillatory regime in addition to previously recognized multimodal stationary phases. We show how local winding numbers predict the steady-state locations in the single mode and bimodal phases, and a flux analysis predicts the respective strengths of the steady-state peaks.
\end{abstract}

\maketitle

\begin{figure}
    \centering
    \includegraphics{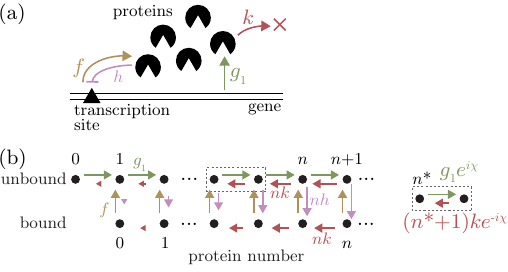}
    \caption{\textbf{Self-repressing gene network.} (a) A gene generates proteins at rate $g_1$. Each protein degrades at rate $k$ or binds to a transcription site at rate $h$, repressing generation when bound. The bound protein unbinds at rate $f$. (b) The network forms a ladder structure defined by protein number $n$ and transcription factor state (bound/unbound). Right inset: to compute {spectral flow,}
     transitions between states $(n^*,1)$ and $(n^*+1, 1)$ are multiplied by phase factors $e^{\pm i\chi}$.}
    \label{fig:model}
\end{figure}

\section{Introduction}

Gene regulatory networks often seem to resemble a giant hairball  of unstructured, heterogeneous, coupled, ultimately stochastic, biochemical reactions \cite{lander2010} characterized by numerous, often poorly known, kinetic parameters \cite{zaghloulsalem2018}. This complexity has evolved so that organisms can cope with a variety of ever-changing environmental challenges. Nevertheless, simple stable collective behaviors often emerge from this complexity. The many stable patterns of gene expression found in different cells are often described as comprising a landscape \cite{goldberg2007}. Abstract, simplified models of stochastic gene networks have been shown to possess attractor landscapes like those of minimally frustrated Hopfield models \cite{sasai2003, zhang2014}. Tripathai, Kessler and Levine have recently analyzed the stability of some specific realistic gene networks to variation of parameters, concluding {that} these networks, in fact, are minimally frustrated {in the same sense} \cite{tripathi2020}.

The quasi-equilibrium landscape picture does not exhaust all the observed regularities of gene regulation, however. The nonequilibrium character of controlled protein synthesis allows for oscillations \cite{potoyan2014} and indeed chaos \cite{aldana2003, heltberg2021}. Such behaviors may be captured through the introduction of gauge fields to the usual reversible dynamics {based} on the gradient of an energy usually envisioned in landscape theory \cite{wang2015, wang2008, zhang2013}. Owing to these emergent gauge fields it is natural to inquire whether topology can give insights into the regularities of gene regulation.

Topology has been found to provide a theoretical prescription for the dimensional reduction of large systems to a lower-dimensional behavior \cite{moore2010, agudo-canalejo2025}. Topological considerations suggest that steady-state response can emerge on the edge of the state space. Crucially, such edge responses, whether as currents or localized states, are robust to random perturbations of the model, an essential element of biological robustness. Topological ideas entered physics in quantum systems \cite{hasan2010, qi2011} but topological thinking has since been developed for other systems including mechanical lattices \cite{kane2014, huber2016, zheng2022}, photonic crystals \cite{lu2014, ozawa2019} and soft matter systems \cite{serra2020, shankar2022, mecke2024}. Topological tools have recently entered biological physics \cite{dasbiswas2018, murugan2017, knebel2020, tang2021, sawada2024a, nelson2024} to describe the circadian rhythm \cite{ zheng2024}, microtubule growth \cite{tang2021}, and chemotaxis adaptation \cite{dasbiswas2018, murugan2017}. Both the existing biological and physical models describe uniform lattices as their state space \cite{dasbiswas2018, murugan2017, knebel2020, tang2021, sawada2024a, nelson2024, agudo-canalejo2025}. The transition network, however, is not uniform in stochastic models of gene networks. In this paper, we develop topological tools for biological networks with heterogeneous transition rates within a lattice configuration and use these tools to describe the simplest self-repressing or self-activating gene switch \cite{hornos2005, schultz2007, choi2010}. 

This simplest gene switch which is turned on or off by a monomeric transcription factor is remarkable in that its deterministic limit suggests that it should not display {multimodal} behavior at all. Nevertheless, the exact solution does display multiple patterns of expression \cite{hornos2005, schultz2007}. {Here, we formally predict these regimes and their phase transitions using topological methods.  We further predict the position of the steady-states of this model in the single and bimodal phases and}
use flux analysis to predict the respective heights of steady-state peaks. 
The complexity of the simplest switch’s behavior comes from near extinction events that occur when the transcription factor number veers on vanishing, at the edge of the state space \cite{schultz2008}. In principle all gene networks have state space with such edges where proteins become extinct at least briefly, so the topological edge features paramount in the simplest gene switch may also manifest themselves in other more realistic situations.

\section{Spectral flow predicts three phases in the gene switch}

Here, we introduce 
{a spectral flow analysis}
that  predicts {several} different dynamical regimes in a biochemical network. We start by analyzing the single transcription factor gene switch \cite{hornos2005, schultz2007}, shown in Fig.~\ref{fig:model}(a). This is the simplest possible switch. In this system, a gene generates proteins at rate $g_1$, while the protein acts as its own repressive transcription factor: it binds to the gene at rate $h$, repressing protein generation. The transcription factor unbinds at rate $f$, restoring generation. Proteins degrade at rate $k$.

This system thus forms a state space described by a vector $p_{ns}$, where $n\in \mathbb{R}_+^0$ is the number of protein molecules in the cell, and $s=0$ and $1$ are the off and on states of the gene, which correspond to the DNA being bound or free of bound  transcription factor (see Fig.~\ref{fig:model}(b)). This probability vector is governed by the master equation
\begin{equation}
    \partial_t p_{ns}=\sum_{mq}\mathcal{W}_{ns,mq}p_{mq},
\end{equation}
where the transition matrix $\mathcal{W}$ specifies the following rates: generation in on state $\mathcal{W}_{(n+1)1, n1}=g_1$, degradation $\mathcal{W}_{ns, (n+1)s}=(n+1)k$, binding $\mathcal{W}_{n0,(n+1)1}=(n+1)h$, and unbinding $\mathcal{W}_{(n+1)1,n0}=f$.
To conserve probability, diagonal terms of the transition matrix $\mathcal{W}$ balance the sum of the outgoing transitions. Due to the state dependence of the degradation and binding rates, this network is nonuniform and thus requires tools beyond those used in the usual quantum context.

\begin{figure}
    \centering
    \includegraphics{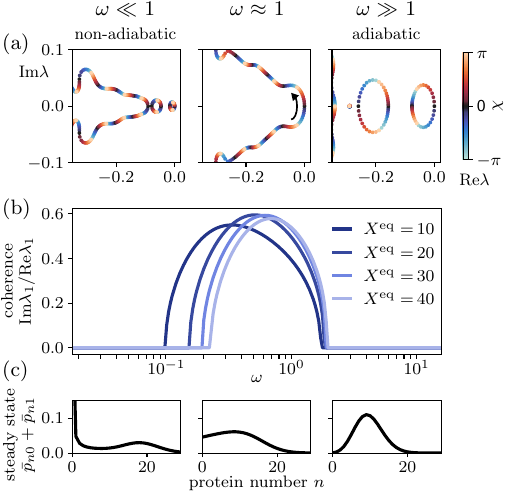}
    \caption{\textbf{{Spectral flow} 
    predicts three distinct phases in a self-repressing gene network.} (a) The network spectrum as a function of $\chi$ shows spectral {flow} in three regimes of adiabaticity $\omega=f/k$, represented by $\omega=0.1$, $1$ and $5$.
    (b) Coherence as a function of adiabaticity $\omega$ reveals an oscillatory regime at intermediate $\omega\approx 1$. (c) Steady state has two peaks in the non-adiabatic phase, spreads over the network in the oscillating phase, and has a single peak in the adiabatic phase. Coherence is shown for $X^\mathrm{ad}=g_1/2k=20$ and varying $X^\mathrm{eq}=f/h$. Panels (a) and (c) use $X^\mathrm{ad}=X^\mathrm{eq}=10$.
    }
    \label{fig:three_regimes}
\end{figure}

To analyze topological properties of the gene network, 
{we introduce a counting field $\chi$, that counts the number of times that a typical trajectory encompasses a chosen transition \cite{sinitsyn2007} (see Appendix~\ref{app:counting_statistics} for details).}
Practically, this entails one rate of the network to be given a complex value whose phase $\chi$ sweeps from $-\pi$ to $\pi$, as illustrated in Fig.~\ref{fig:model}(b). Note that this field
could have been inserted on any rate \rev{along a typical trajectory of the system without changing the counting statistics. In our calculations, we choose a horizontal link at the number of protein molecules $n^*$, at which the vertical current vanishes. This ensures that the link stays on the typical trajectory, independent of the system parameters.} Introducing $\chi$ allows us to count and envision possible cyclic paths through the state space. 
Counting statistics alone, however, do not distinguish between different phases of the network.

Hence, we additionally plot the spectral flow, i.e.~the spectrum along the full domain of $\chi$ from $-\pi$ to $\pi$---a method used to probe the system global winding (interpreted in quantum systems as insertion of a magnetic flux) \cite{gong2018}.
While the spectrum of the original  transition matrix  $\mathcal{W}$ is {shown in black} 
in Fig.~\ref{fig:three_regimes}(a), we see that sweeping the phase $\chi$ from $-\pi$ to $\pi$ produces a continuous change in the spectrum that forms loops, where the color corresponds to different values of $\chi$ {(see Appendix~\ref{app:global_spectrum} for details about calculation)}.

\begin{figure*}
    \centering
    \includegraphics{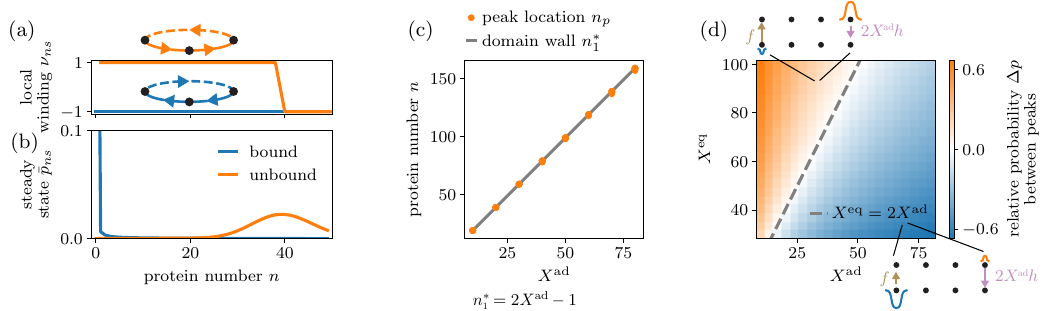}
    \caption{\textbf{In the non-adiabatic phase, local winding of bound and unbound chains individually predict two peaks.} (a) Local winding number is tracked along the bound (blue) and unbound (orange) chains. It is computed under periodic boundary conditions, as illustrated in two cartoons. (b)  Steady-state peaks occur at an edge (bound) and a domain wall (unbound), as predicted by local winding numbers. (c) Observed peak locations in the unbound states $n_p$ (orange dots) match predicted domain wall locations $n^*_1$ (gray line) for a range of parameter values. (d) Relative probability between peaks $\Delta p=\sum_n p_{n1}-p_{n0}$ for a range of $X^\mathrm{ad}$ and $X^\mathrm{eq}$; peaks are equal when $X^\mathrm{eq}=2X^\mathrm{ad}$. Insets: right peak dominates when $X^\mathrm{eq}>2X^\mathrm{ad}$, and is smaller when $X^\mathrm{eq}<2X^\mathrm{ad}$. 
    Plots use $X^\mathrm{eq}=X^\mathrm{ad}=20$ and $\omega=0.01$.}
    \label{fig:nonadiabatic}
\end{figure*}

Notably, the loops appear qualitatively different for different values of the adiabaticity parameter $\omega=f/k$.
{Adiabaticity, as introduced in Ref.~\cite{walczak2005} describes whether the DNA site has equilibrated between the bound and unbound states or not, corresponding to the adiabatic ($\omega\gg1$) regime where local equilibrium is achieved or to the non-adiabatic ($\omega\ll1$) regime, when it is not.} 
The network also depends on additional parameters $X^\mathrm{ad}=g_1/2k$ and $X^\mathrm{eq}=f/h$. {We further globally modulate these parameters to demonstrate the stability of our topological predictions to variations in transition rates \cite{huang2017}. Note that introducing random variations of individual transitions (rather than global rate constants), as is commonly done in topological stability analysis \cite{murugan2017, zheng2024}, is rather unnatural for biochemical gene networks, as this would imply randomness in rates for different numbers of protein molecules.}  

{In Fig.~\ref{fig:three_regimes}(a),} for both $\omega\ll1$ and $\omega\gg1$, small spectral loops appear near the steady state that are composed of only two states that fold back on themselves. The next two states away from the steady state form another separate loop. In contrast, for $\omega\approx 1$, all the states connect to form a continuous large loop. These differences in loop size point to distinct physics, since stochastic transitions between only two states always have zero oscillatory coherence \cite{barato2017} in contrast to when multiple states participate. 

To test if indeed we obtain regimes with different oscillatory features, we analyze the coherence of the system. For this, we calculate the ratio of the imaginary and real parts of the first non-trivial eigenvalue of the transition matrix $\mathcal{W}$, $\mathcal{R}=\Im\lambda_1/\Re\lambda_1$. 
This expression gives the number of coherent oscillations weighted by their lifetime \cite{barato2017}. Indeed, we find that oscillations emerge at $\omega\approx 1$ as indicated by a non-zero coherence in Fig.~\ref{fig:three_regimes}(b), for several values of parameter $X^\mathrm{eq}$. These oscillations were not noticed in  previous studies of this gene network \cite{hornos2005, schultz2007}, and were only revealed upon analysis of this {spectral flow.}
Instead, the previous studies described only a regime of bistability for $\omega\ll1$ and of monostability for $\omega\gg1$, as we can see in plots of the steady state in Fig.~\ref{fig:three_regimes}(c). While the steady state of $\omega\approx 1$ appears flat and featureless, this is nevertheless consistent with the system cycling across the whole network in this explicitly non-equilibrium regime. 

Crucially, as we go between the small and large loop regimes, the system goes through a non-reciprocal phase transition that is characterized by a pitchfork bifurcation of the eigenvalues of the complex master equation \cite{vansaarloos2024}. Specifically, the two eigenvalues closest to the steady state coalesce to form an exceptional point and then split into a pair of complex-conjugated eigenvalues (see Appendix~\ref{app:exceptional_point} for details). This marks a transition between a chiral and a stationary phase \cite{fruchart2021}, which has also been observed in other platforms, such as active matter \cite{fruchart2021} or optical quantum gases \cite{ozturk2021}.

The global winding number \cite{sawada2024a}{, determined from the spectral flow as} 
\begin{equation}
    \nu=\sum_{\varepsilon=\pm 0}\frac{1}{2\pi i}\int\limits_0^{2\pi} d\chi \partial_\chi \log\det \mathcal{W}(\chi+i \varepsilon),
    \label{eq:winding}
\end{equation}
turns out to be $+1$ for all three regimes. When the network ends in a sink, i.e. extinction in the regime of death with strictly no regeneration, the winding number becomes 0. In this paper, we focus on ergodic networks that do regenerate, albeit at a small rate. 

\begin{figure*}
    \centering
    \includegraphics{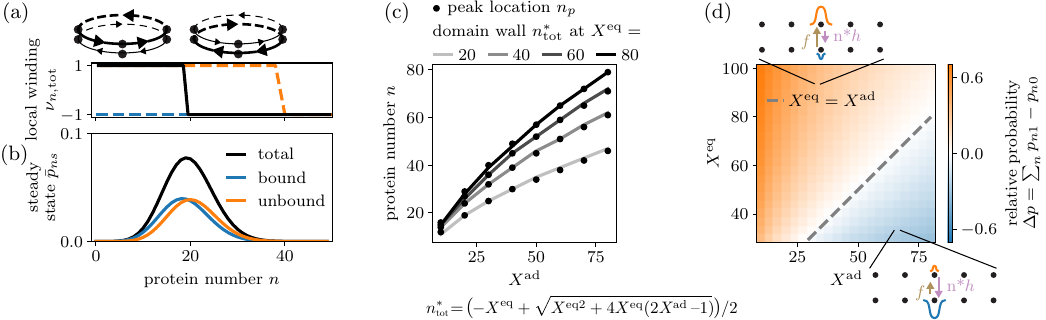}
    \caption{\textbf{In the adiabatic phase, combined local winding determines domain wall and peak location.} (a) Local winding number is tracked along two averaged chains (black). Cartoons illustrate networks with periodic boundary conditions, where either the unbound or bound chain dominates the net winding direction. (b) Steady state peak in the total probability (black) is determined by the domain wall in the local winding number. (c) Observed peak locations $n_p$ (black dots) match predicted domain wall locations $n^*_\mathrm{tot}$ (gray lines) for a range of parameter values. (d) Relative probability between unbound and bound chains $\Delta p=\sum_n p_{n1}-p_{n0}$ for a range of $X^\mathrm{ad}$ and $X^\mathrm{eq}$. probabilities are equal when $X^\mathrm{eq}=X^\mathrm{ad}$. Insets: unbound probability dominates when $X^\mathrm{eq}>X^\mathrm{ad}$, and is smaller when $X^\mathrm{eq}<X^\mathrm{ad}$.
    Plots use $X^\mathrm{eq}=X^\mathrm{ad}=20$ and $\omega=10$.}
    \label{fig:adiabatic}
\end{figure*}

\section{In the non-adiabatic phase, local winding number predicts {bimodality} and location of steady state peaks}

While the {spectral flow}
can distinguish between different dynamical regimes and identifies the phase transitions, it would be useful to predict specific quantities of interest within each phase, such as the location of the steady-state peak and the relative heights of different peaks. Here, we introduce a \emph{local} winding number for this purpose. 
A winding number in a single one-dimensional chain predicts the accumulation of probability density at the system edge {or domain wall}, also known as the non-Hermitian skin effect \cite{okuma2020, borgnia2020, sawada2024a, sawada2024b}.  However, these winding numbers are usually calculated under periodic boundary conditions \cite{sawada2024a, sawada2024b}. The absence of translation symmetry in {the transition rates as a function of protein number}
in many networks including ours, renders such a global approach inapplicable. {Moreover, the location of a domain wall in these genetic networks  is not determined through an abrupt change in transition rates, as happens at the edge of uniform systems.} \rev{In Hermitian systems, these challenges have been solved using local topological invariants, such as as Chern markers \cite{bianco2011}, which are computed in real space and insensitive to spatial disorder. However, gene networks are explicitly non-Hermitian and therefore need a non-Hermitian analogue of such local topological invariants.}

To address these challenges in  inhomogeneous \rev{non-Hermitian} networks, we develop a \emph{local} winding number that is evaluated {independently} at each protein number $n$ of the network. In the non-adiabatic phase $\omega\ll 1$, slow transitions between the bound and unbound chains allow us to view the gene switch as two separate chains and define their 
{local winding numbers} independently. For each chain, we choose a unit cell that consists of the $n$-th state of the chain together with transitions connecting it to the right neighbor.  Repeating this unit cell many times leads to 
{a new periodic system with transition matrix 
\begin{equation}
    \mathcal{W}^\mathrm{n0}(\chi)=(n+1)k (e^{-i\chi}-1)
\end{equation} 
for the $n$-th unit cell of the bound chain, and 
\begin{equation}
    \mathcal{W}^\mathrm{n1}(\chi)= g_1(e^{i\chi}-1)+(n+1)k (e^{-i\chi}-1)
\end{equation}
for the $n$-th unit cell of the unbound chain, with $\chi\in[-\pi,\pi]$.}

We compute the {$n$-th} local winding {number} in both chains by plugging these transition matrices in Eq.~\eqref{eq:winding},
to show in Fig.~\ref{fig:nonadiabatic}(a) how they change with the protein number $n$ along the network. In the bound chain, the local winding is $\nu_\mathrm{n0}=-1$ along the entire chain, leading to the steady state localized on the left edge at $n^*_0=0$ (blue in Fig.~\ref{fig:nonadiabatic}(b)). In the unbound chain, the local winding number switches from $\nu_\mathrm{n1}=+1$ to $\nu_\mathrm{n1}=-1$ at the protein number $n^*_1=g_1/k-1=2X^\mathrm{ad}-1$, where the forward and backward transitions become equal to each other. This signals a domain wall at $n^*_1$, where we also find the steady state peak; see the orange line in Fig.~\ref{fig:nonadiabatic}(b) for $X^\mathrm{eq}=X^\mathrm{ad}=20$. We verify that our analytical prediction for the domain wall matches the location of the steady state peak, evaluated numerically for a range of parameters $X^\mathrm{ad}$ and $X^\mathrm{eq}$ (see Fig.~\ref{fig:nonadiabatic}(c)). 

We can further predict which of the two peaks, bound or unbound, will be larger. We use a principle of flux balance similar to Kirchhoff's laws for electrical circuits, derived in Appendix~\ref{app:current_balance}. This principle states that the total upward flux in the gene network must be equal to the total downward flux such that $\sum_n J_{(n+1)1,n0}=0$. Here, the probability flux between the $i$-th and $j$-th states is $J_{ji}=\mathcal{W}_{ji}p_i-\mathcal{W}_{ij}p_j$. 
In the non-adiabatic phase, there are two narrow peaks at the steady state. 
The bound peak at $n=0$ with the total probability $p_0$ causes an upwards flux $J_\uparrow\approx fp_0$.
At the same time, the unbound peak at $n\approx2X^\mathrm{ad}$ with the total probability $p_1$ causes a downwards flux $J_\downarrow\approx 2X^\mathrm{ad}hp_1$. Since $J_\uparrow=J_\downarrow$, the two peaks are equal when $2X^\mathrm{ad}=f/h=X^\mathrm{eq}$ (dashed line in Fig.~\ref{fig:nonadiabatic}(d)). We predict that the unbound peak is larger than the bound peak when $X^\mathrm{eq}>2X^\mathrm{ad}$ and vice versa, which we also confirm numerically in Fig.~\ref{fig:nonadiabatic}(d).

\section{In the adiabatic phase, the local winding number predicts the location of the steady-state peak}

In the adiabatic phase $\omega\gg1$, fast transitions between the bound and unbound chains cause the system to average over these states, similar to what happens in the Shea-Ackers deterministic model \cite{shea1985}. Due to this averaging, we define a combined local winding number for coupled bound and unbound states. This local winding would then predict the steady state of the full gene network.

To define this invariant, we choose a unit cell that consists of the $n$-th state in the bound chain and the $(n+1)$-th state in the unbound chain, together with transitions between these states as well as transitions to and from their right neighbors. Imposing periodic boundary conditions on this unit cell, we obtain the combined local periodic network with transition matrix 
\begin{widetext}
\begin{equation}
    \mathcal{W}^{n,\mathrm{tot}}(\chi)=\begin{pmatrix}
        (n+1)k (e^{-i\chi}-1) - f & (n+1)h \\
        f & g_1(e^{i\chi}-1)+(n+2)k (e^{-i\chi}-1) - (n+1)h
    \end{pmatrix}.
    \label{eq:combined_matrix}
\end{equation}
\end{widetext}
We compute the corresponding local winding number using Eq.~\eqref{eq:winding}. Plotting the combined local winding as a function of the protein number $n$ in Fig.~\ref{fig:adiabatic}(a), we observe that it changes sign across a domain wall, when the winding of the bound chain starts dominating over the unbound chain, as illustrated in cartoons above the plot.

The domain wall in the combined winding number determines the location of the steady state peak, as we illustrate in Fig.~\ref{fig:adiabatic}(b).
To analytically derive the domain wall location, we use the fact that the sign of the winding number is the same as the sign of the probability flux along a periodic network \cite{sawada2024a}.
Therefore, the local winding number changes sign where probability flux vanishes. We use this to infer that the domain wall (details in Appendix~\ref{app:domain_wall}) will be located at
\begin{equation}
    n^*_\mathrm{tot}= \left(-X^\mathrm{eq}+\sqrt{(X^\mathrm{eq})^2+4X^\mathrm{eq}(2X^\mathrm{ad}-1)}\right)/2.
\end{equation}
To confirm this prediction, we verify in Fig.~\ref{fig:adiabatic}(c) that the domain wall determines the peak location for a range of parameters $X^\mathrm{ad}$ and $X^\mathrm{eq}$.
We note that when defining the {spectral flow}
{in Fig.~\ref{fig:three_regimes}(a)},
we used the same value of protein number $n^*=n^*_\mathrm{tot}$ to introduce the complex phase factor.

Similar to what was done for the non-adiabatic phase, we use flux balance to predict the probability distribution between bound and unbound chains. With both bound and unbound peaks located at $n=n^*_\mathrm{tot}$, the up and down fluxes are $J_\uparrow\approx fp_0$ and $J_\downarrow\approx n^*_\mathrm{tot}hp_1$. Probabilities of being bound or unbound are equal when $n^*_\mathrm{tot}=f/h=X^\mathrm{eq}$, which is true when $X^\mathrm{ad}\approx X^\mathrm{eq}$ (dashed line in Fig.~\ref{fig:nonadiabatic}(d)). We numerically confirm this prediction in Fig.~\ref{fig:nonadiabatic}(d).

\section{Conclusion}

We have developed topological methods for stochastic systems that have inhomogeneous transition rates that break translation symmetry in number space and demonstrate this approach on the simplest gene switch network. Using these methods, the analysis predicts three different phases even for this simple network, and several crucial properties of its steady state, such as peak location and relative heights. These results pave the way for future explorations using the lens of topology. For instance, relaxation times {have been predicted to scale differently with system size depending on} topology \cite{sawada2024a}, which would have biological consequences. Another interesting regime is when the global winding number goes to zero, i.e. in the non-ergodic case \cite{sawada2024b} which applies where the transcription factor can go completely extinct without the possibility of regeneration. 

{In the future,} our methods can be generalized to other biological networks with similar underlying structure. This includes the case of dimer binding instead of monomer binding where even the deterministic treatment gives multiple stable states \cite{schultz2008, amoutzias2008, smale2012, feng2011}, or going to higher-dimensional networks, such as those consisting of multiple genes or proteins \cite{schultz2008, potoyan2014, wang2016}. {It would also be interesting to use topological ideas for deterministically} oscillatory gene networks such NF$\kappa$B/I$\kappa$B  \cite{potoyan2014, wang2016}. 
In cases where transcription factors can go strictly extinct we see that many gene networks have the possibility of having an unavoidable death. It is an interesting question whether real biological networks have evolved to avoid such catastrophes through their topology. 

\section*{Acknowledgments}
We thank Dexin Li and Erwin Frey for helpful insights into winding numbers and flux balance respectively. This work was supported by the NSF Center for Theoretical Biological Physics (PHY-2019745) and the NSF CAREER Award (DMR-2238667). 

\appendix

\section{Global spectrum}

\label{app:global_spectrum}
The gene switch is determined by a transition matrix $\mathcal{W}$ introduced in the main text. In this Appendix, we explain how we computed its spectral flow.

First, we take one link of the gene network $n^*\rightleftharpoons n^*+1$, and allow it to take on a complex value whose phase $\chi$ can be swept from $-\pi$ to $\pi$, $\mathcal{W}_{(n^*+1)1,n^*1}(\chi)=\mathcal{W}_{(n^*+1)1,n^*1}(0)e^{i\chi}$ and $\mathcal{W}_{n^*1,(n^*+1)1}(\chi)=\mathcal{W}_{n^*1,(n^*+1)1}(0)e^{-i\chi}$. Note that this phase could be inserted on any rate. In our calculation, we choose $$n^*=\left(-X^\mathrm{eq}+\sqrt{(X^\mathrm{eq})^2+4X^\mathrm{eq}(2X^\mathrm{ad}-1)}\right)/2,$$ which corresponds to the domain wall derived in Appendix~\ref{app:domain_wall}.

Then, we compute the spectrum of the periodic matrix $\mathcal{W}(\chi)$ using exact diagonalization for a range of $\chi\in[-\pi, \pi]$. In our calculation, the network consists of 60 states, 30 in each chain, bound and unbound. Each transition matrix is calculated for 21 value of $\chi$ equidistantly distributed between $-\pi$ and $\pi$ values. Eigenvalues of the transition matrix $\mathcal{W}(\chi)$ flow continuously as the phase $\chi$ is swept between $-\pi$ and $\pi$, thus forming a spectral flow.

\begin{figure*}
    \centering
    \includegraphics{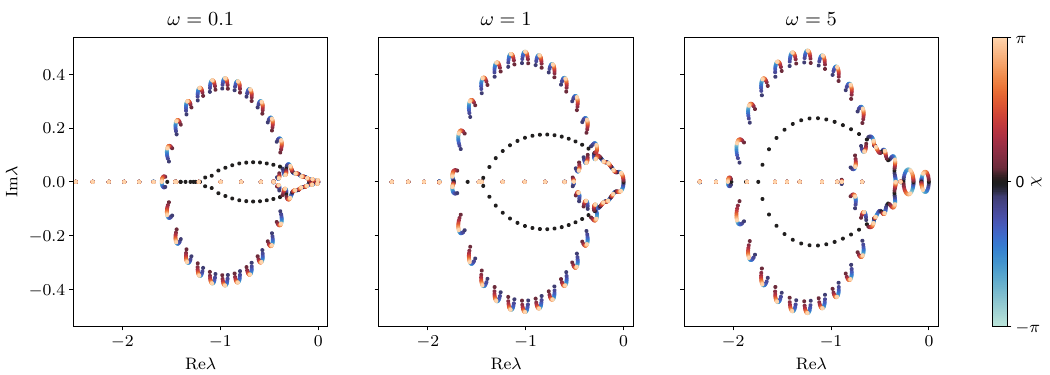}
    \caption{Spectrum of the transition matrix $\mathcal{W}(\chi)$ with manually imposed periodic boundary conditions for a range of $\chi\in[-\pi, \pi]$ and three values of adiabaticity $\omega$. Other system parameters are set to $X^\mathrm{ad}=X^\mathrm{eq}=10$.}
    \label{fig:spectrum}
\end{figure*}

Note that the spectra in Fig.~2(a) of the main text are zoomed in to distinguish fine details near the steady state, so not all calculated eigenvalues are visible in the figure. We plot the zoomed-out spectra in Fig.~\ref{fig:spectrum} for the same set of parameters.

\section{Counting statistics interpretation of the field $\chi$}

\label{app:counting_statistics}
In this Appendix we provide a physical interpretation of the spectral flow introduced in Appendix~\ref{app:global_spectrum}. It can be understood through the number of times that a typical trajectory crosses the link with the phase $\chi$, the so called counting statistics introduced in Ref.~\cite{sinitsyn2007}.To count the number of trajectory crossings, this formalism introduces a probability generating function 
$$Z(\chi)=\langle1|e^{\mathcal{W}(\chi)T}|\bar{p}\rangle,$$ 
where $T$ is time over which we observe the system, $\chi$ has the meaning of the counting field, and $\log Z(\chi)$ is the full counting statistics. This means that the derivative $$\frac{d}{d(i\chi)}\log Z(\chi)\big|_{\chi=0}$$ gives the number of times that a typical trajectory crosses the chosen link over the time $T$. In our system, this corresponds to the number of clockwise loops that the system makes around the network. The corresponding current can be computed as the derivative of the first eigenvalue of the transition matrix \cite{sawada2024a}: $$J=\frac{d}{d(i\chi)} \lambda_0(\chi)\big|_{\chi=0}.$$

\begin{figure*}[h]
    \centering
    \includegraphics{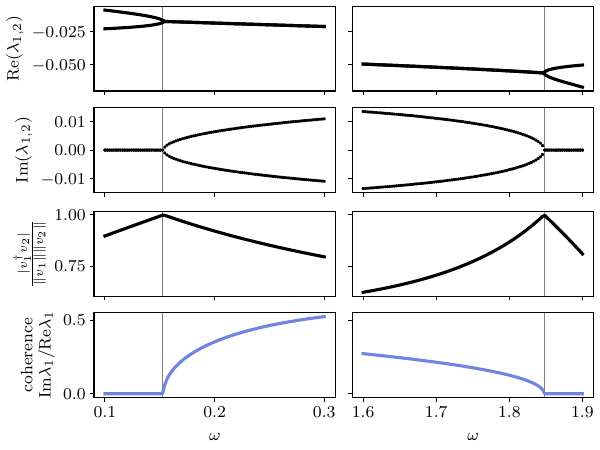}
    \caption{Exceptional point phase transition in the gene switch. We plot two closest to the steady state eigenvalues, and a scalar product of the corresponding eigenvectors as a function of adiabaticity $\omega$. We plot coherence of oscillations for the same parameter range. These plots are made for parameter values $X^\mathrm{eq}=X^\mathrm{ad}=20$.}
    \label{fig:exceptional_point}
\end{figure*}

Finally, we would like to compare this interpretation with a flux insertion method used to compute spectral flows in disordered quantum systems \cite{gong2018}. There, the phase $\chi$ is usually distributed evenly across all transitions of the entire network. In this case, phase $\chi$ can be interpreted as a magnetic flux inserted in the middle of a closed loop, affecting the phase of transition amplitudes. While the physical interpretation in quantum and stochastic systems is different, both methods enable computing spectral flows and studying topological properties of these systems.\\

\section{Exceptional point phase transition}

\label{app:exceptional_point}

In the main text, we discussed how three regimes of the gene switch are separated by exceptional point phase transitions \cite{fruchart2021, vansaarloos2024}. Here, we provide a more detailed analysis of these phase transitions. 

Exceptional points are characterized by a simultaneous coalescence of eigenvalues and eigenvectors of a non-Hermitian matrix. We demonstrate that the two closest to the steady state eigenvalues of the transition matrix $\mathcal{W}$ go through an exceptional point. For this we plot in Fig.~\ref{fig:exceptional_point} how their real and imaginary parts change with adiabaticity $\omega=f/k$. We also plot a normalized scalar product of the corresponding eigenvectors, which turn to 1 when the two vectors become linearly dependent. This happens at the same values of $\omega$ at which the eigenvalues coalesce, proving that the system goes through exceptional points phase transitions.

Plotting in Fig.~\ref{fig:exceptional_point} the coherence of gene network oscillations for the same parameter range, we confirm that coherence changes from zero to non-zero values at exceptional point phase transitions.

\section{Current balance follows from Kirchhoff's laws}

\label{app:current_balance}

In this Appendix, we derive the current balance principle, which was used in the main text, from the Kirchhoff's laws. To start, we define the probability current from the $i$-th to the $j$-th state as
\begin{equation}
    J_{ji}=\mathcal{W}_{ji}p_i-\mathcal{W}_{ij}p_j.
\end{equation}
Currents in the gene network must obey the Kirchhoff's laws:
\begin{align}
    J_{(n+1)0,n0}-J_{n0,(n-1)0}&=-J_{(n+1)1,n0}, \label{eq:Kirchhoff_lower}\\
    J_{(n+2)1,(n+1)1}-J_{(n+1)1,n1}&=J_{(n+1)1,n0},
\end{align}
while at the left edge the Kirchhoff's laws are fulfilled by
\begin{equation}
    -J_{10,00}=J_{11,00}=J_{21,11}.
    \label{eq:kirchhoff_left}
\end{equation}
Summing up Eq.~\eqref{eq:Kirchhoff_lower} for $n$ from $1$ to $\infty$, and taking into account Eq.~\eqref{eq:kirchhoff_left}, we derive that the vertical current summed over the network is zero
\begin{equation}
    \sum_{n=0}^\infty J_{(n+1)1,n0}=0.
\end{equation}

\section{Analytical derivation of the domain wall in the adiabatic phase}
\label{app:domain_wall}

In this Appendix, we analytically derive the domain wall location for the combined local winding number in the adiabatic phase. For this, we use the fact that the winding number changes its sign when currents in the corresponding periodic network vanish \cite{sawada2024a}. To find this location, we track how vertical and horizontal currents in a local periodic network given by Eq.~\eqref{eq:combined_matrix} change as a function of protein number $n$.
Assuming that the total probability in a local network splits into probability $p_0$ in the lower chain and $p_1$ in the upper chain, we can express the currents in the $(n-1)$-th local network as
\begin{align}
    J_\mathrm{vertical}&=fp_0-nhp_1, \label{eq:vert_current}\\
    J_\mathrm{horizontal}&=J_0+J_1=-nkp_0+(g_1-(n+1)k)p_1, \label{eq:horiz_current}
\end{align}
where the horizontal current has two contributions, from the lower and the upper chains.
To find the domain wall location, we set both vertical and horizontal currents to zero. From $J_\mathrm{vertical}=0$ we get 
\begin{equation}
    p_0=nhp_1/f.
\end{equation}
Plugging this into $J_\mathrm{horizontal}=0$, we get
\begin{equation}
    n^2+(f/h)n-(f/h)(g_1/k-1)=0.
\end{equation}
The positive solution of this equation, parameterized by $X^\mathrm{ad}=g_1/2k$ and $X^\mathrm{eq}=f/h$, is
\begin{equation}
    n^*_\mathrm{tot}= \left(-X^\mathrm{eq}+\sqrt{(X^\mathrm{eq})^2+4X^\mathrm{eq}(2X^\mathrm{ad}-1)}\right)/2.
\end{equation}
This protein number corresponds to the domain wall of the local winding number.

\bibliographystyle{ieeetr}
\bibliography{references}

\end{document}